\PassOptionsToClass{10pt}{revtex4-2}
\documentclass[aps,prl,showpacs,floatfix,twocolumn,byrevtex,superscriptaddress]{revtex4-1}
\usepackage{amsmath}
\usepackage{amssymb}
\usepackage{amstext}
\usepackage{amsopn}
\usepackage{amsfonts}
\usepackage{amsxtra}
\usepackage[english]{babel}
\usepackage{graphicx}
\usepackage{bm}
\usepackage{multirow}
\usepackage{dcolumn}
\usepackage{color}
\usepackage{hyperref}
\usepackage{todonotes}
\usepackage{verbatim}
\usepackage{soul}
\usepackage{xcolor}
\usepackage{lineno}
\usepackage{siunitx}
\usepackage[version=3]{mhchem}
  \usepackage{natbib}

\def\NIS{Ni$_{3}$In$_{2}$S$_{2}$}
\def\CSS{Co$_{3}$Sn$_{2}$S$_{2}$}
\def \kag{Kagome}
\def \kag{Kagome}

\begin{document}

\title{Endless Dirac nodal lines in \kag{}-metal \NIS{}}

\author{Tiantian Zhang} \email{zhang.t.ac@m.titech.ac.jp}
\affiliation{Department of Physics, Tokyo Institute of Technology, Okayama, Meguro-ku, Tokyo 152-8551, Japan}
\affiliation{Tokodai Institute for Element Strategy, Tokyo Institute of Technology, Nagatsuta, Midori-ku, Yokohama, kanagawa 226-8503, Japan}
\author{Turgut Yilmaz Yilmaz}
\affiliation{National Synchrotron Light Source II, Brookhaven National Laboratory, Upton, New York 11973, USA}
\author{Elio Vescovo}
\affiliation{National Synchrotron Light Source II, Brookhaven National Laboratory, Upton, New York 11973, USA}
\author{Haoxiang Li}
\affiliation{Material Science and Technology Division, Oak Ridge National Laboratory, Oak Ridge, Tennessee 37831, USA}
\author{Rob G. Moore}
\affiliation{Material Science and Technology Division, Oak Ridge National Laboratory, Oak Ridge, Tennessee 37831, USA}
\author{Ho Nyung Lee}
\affiliation{Material Science and Technology Division, Oak Ridge National Laboratory, Oak Ridge, Tennessee 37831, USA}
\author{Hu Miao} \email{miaoh@ornl.gov}
\affiliation{Material Science and Technology Division, Oak Ridge National Laboratory, Oak Ridge, Tennessee 37831, USA}
\author{Shuichi Murakami}
\affiliation{Department of Physics, Tokyo Institute of Technology, Okayama, Meguro-ku, Tokyo 152-8551, Japan}
\affiliation{Tokodai Institute for Element Strategy, Tokyo Institute of Technology, Nagatsuta, Midori-ku, Yokohama, kanagawa 226-8503, Japan}
\author{Micheal A. McGuire} \email{mcguirema@ornl.gov}
\affiliation{Material Science and Technology Division, Oak Ridge National Laboratory, Oak Ridge, Tennessee 37831, USA}

\date{\today}

\begin{abstract}
 
Topological semimetals are a frontier of quantum materials. In multi-band electronic systems, topological band-crossings can form closed curves, known as nodal lines. In the presence of spin-orbit coupling and/or symmetry-breaking operations, topological nodal lines can break into Dirac/Weyl nodes and give rise to transport properties, such as the chiral anomaly and giant anomalous Hall effect. Recently the time-reversal symmetry-breaking induced Weyl fermions are observed in a \kag{}-metal \CSS{}, triggering interests in nodal-line excitations in multiband \kag{} systems. Here, using first-principles calculations and symmetry based indicator theories, we find six endless nodal lines along the stacking direction of \kag{} layers and two nodal rings in the \kag{} plane in nonmagnetic Ni$_{3}$In$_{2}$S$_{2}$. The linear dipsersive electronic structure, confirmed by angle-resolved photoemission spectroscopy, induces large magnetoresistance up to 2000\% at 9~T. Our results establish a diverse topological landscape of multi-band \kag{} metals.

\end{abstract}
\maketitle

\section*{Introduction}
\begin{figure*}
\includegraphics[scale=0.35]{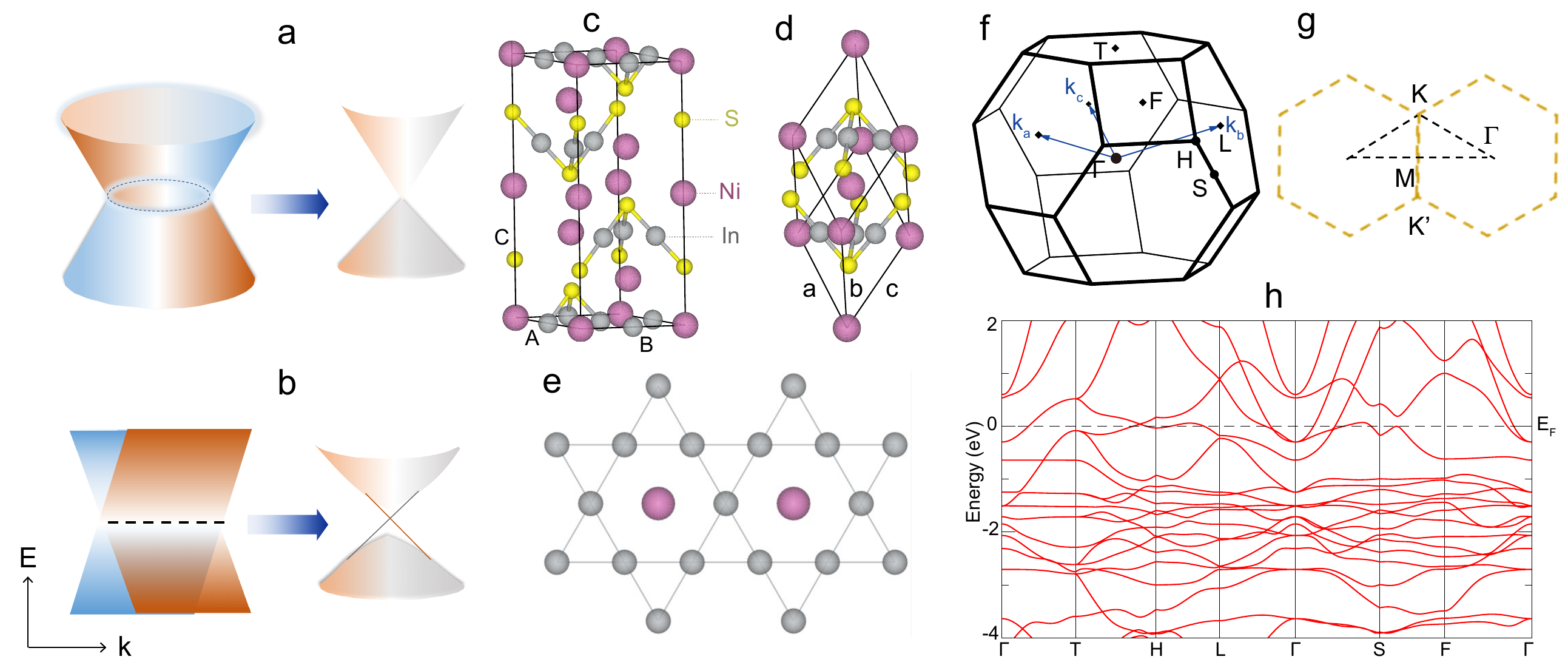}
\caption{\textbf{Crystal structure and electronic band structure for \NIS.} \textbf{a} Weyl/Dirac points that can be obtained from nodal ring band crossings in the Kagome lattice by tuning parameters like magnetism, spin-orbit coupling and breaking crystalline symmetries. \textbf{b} Topological (crystalline) insulator states obtained from nodal line band crossing after considering spin-orbit coupling. \textbf{c-d} Conventional cell and primitive cell for the Kagome-lattice material Ni$_{3}$In$_{2}$S$_{2}$. \textbf{e} Kagome layer in Ni$_{3}$In$_{2}$S$_{2}$. \textbf{f-g} Three-dimensional BZ and surface BZ along $z$ direction for Ni$_{3}$In$_{2}$S$_{2}$. Throughout the manuscript, we use the primitive cell and the 3D BZ for theoretical calculations. \textbf{h} Spinless electronic band structure of Ni$_{3}$In$_{2}$S$_{2}$ along some high-symmetry lines, with band crossings near the Fermi level highlighted by dashed circles.} 
\label{fig:FIG1}
\end{figure*}

\kag{} metals, consisting of a geometrically frustrated \kag{} sublattice, are fertile platforms for emergent topological states such as quantum spin liquids and quantum Hall states \cite{Tang2011,Sheng2011,Neupert2011,Sun2011,Matsumoto2018, Ye2018, Yin2018, LiuDF2019, Yin2020, Liu2020CoSn, Meier2020, Kang2020CoSn, Kang2020FeSn, Sales2021, Ortiz2019, LiM2021, LiHX2021,belopolski2021signatures}. Theoretically, electrons residing on the corner-shared triangle network create non-trivial quantum interference among the three sub-lattices and gives rise to flat bands, saddle points and Dirac fermions \cite{Wang2013,Thomale2013,Tang2011,Sheng2011,Neupert2011,Sun2011, Ye2018}. Studies of \kag{} metals have been focusing on these characteristic excitations; however, in the presence of multi-bands near the Fermi level, topological nodal lines can emerge and serve as an avenue to realize topological semimetals and insulators~\cite{burkov2011topological,fang2016topological}. A widely studied example is the {ferromagnetic} shandite \kag{}-metal Co$_3$Sn$_2$S$_2$, where the mirror symmetry protected nodal ring (Fig.~\ref{fig:FIG1} a) breaks into Weyl nodes by {its strong spin-orbit coupling} \cite{xu2018topological,LiuDF2019,morali2019fermi,jiao2019signatures,shen2019anisotropies,chen2019pressure,ding2019intrinsic,xu2020electronic,li2020epitaxial,tanaka2020topological,gopal2020observation,muechler2020emerging,guin2019zero,belopolski2021signatures}. Here we show that the shandite \kag{}-metals can host another nodal line, an endless Dirac-nodal line (Fig.~\ref{fig:FIG1} b), in their electronic structure. Using density functional theory (DFT) calculations and symmetry-based indicator theories, we demonstrate that the non-magnetic Ni$_{3}$In$_{2}$S$_{2}$ is a topological semimetal with six endless Dirac-nodal lines near the Fermi level. The linearly dispersive band structure yields small effective mass and high mobility of the conduction electrons, which are responsible for a large magnetoresistance (MR) up to 2000\% at 9~T. Our results uncover the diverse landscape of multiband \kag{}-metals and suggest that the hole-doped \NIS{} as a Dirac/Weyl semimetal platform for quantum effects.

\section*{Results}
\subsection*{Crystal structure, electronic bands and topology of \NIS}
    The nonmagnetic shandite \kag{}-metal belongs to space group No.166 ($\mathrm{R}\bar{3}\mathrm{m}$). The conventional cell and primitive cell of \NIS{} are shown in Fig.~\ref{fig:FIG1} c and d, respectively. Ni-atoms form a 2D \kag{} sublattice within the In-Ni layers, as shown in Fig. ~\ref{fig:FIG1} e. Figure~\ref{fig:FIG1} f and g show the three-dimensional Brillouin zone (BZ) of the primitive cell and two-dimensional surface BZ in the conventional cell, respectively. Throughout the manuscript, we use the primitive cell and the three-dimensional BZ for theoretical calculations, and the surface BZ for the angle-resolved photoemission spectroscopy (ARPES) measurement. 

\begin{figure}
\includegraphics[scale=1]{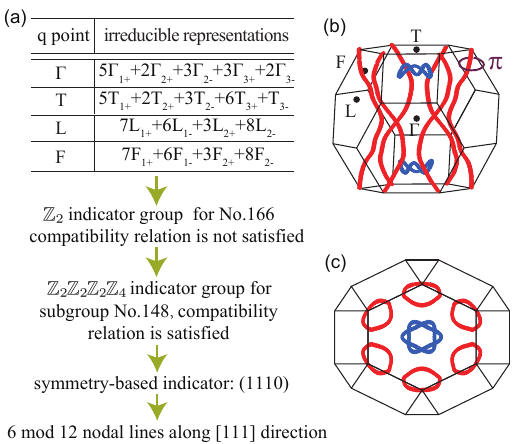}\caption{\textbf{Topological diagnosis for the nodal lines in \NIS{}.} \textbf{a} Diagnostic process for topological nodal lines in Ni$_{3}$In$_{2}$S$_{2}$ by symmetry-based indicator theory and beyond (see text for more details). \textbf{b} Distributions of six endless node-line (red) and two node-ring (blue) band crossings in the first BZ, carrying $\pi$ Berry phase. \textbf{c} Top view of the nodal lines and nodal rings in the BZ. 
\label{fig:FIG3}}
\end{figure}

    Figure~\ref{fig:FIG1} h shows the DFT calculated band structure without spin-orbit coupling (SOC), {since the SOC is small compared to the experimental resolution (Details are in the Supplementary Note. 3)}. Two band crossings {(highlighted by dashed circles)} along $\mathrm{T}$-$\mathrm{H}$ and $\Gamma$-$\mathrm{S}$ high-symmetry lines are identified near the Fermi level, $\textrm{E}_\textrm{{F}}$. {These crossings are composed mainly by the Ni 3$d$-orbitals (see Supplementary Figure. 11)}. Since Ni$_{3}$In$_{2}$S$_{2}$ preserves both time-reversal, $\mathcal{T}$, and inversion, $\mathcal{P}$, symmetries, these linear band crossings may belong to nodal lines/rings. To determine the topology of these linear crossings, we use symmetry-based indicator theories~\cite{song2018diagnosis,zhang2020diagnosis,zhang2021predicting} for further diagnosis. Figure~\ref{fig:FIG3} summarizes the diagnostic processes. Under the compatibility condition, we find the space group No.~148 is the maximum subgroup, having a nontrivial symmetry-based indicator group of $\mathbb{Z}_{2}\mathbb{Z}_{2}\mathbb{Z}_{2}\mathbb{Z}_{4}$, which is used to determine the complete topological information for the band crossings near the Fermi level~\cite{zhang2020diagnosis,zhang2021predicting,kruthoff2017topological,po2017symmetry,bradlyn2017topological,song2018diagnosis}. Topological invariants for the symmetry-based indicator formulas of the space group No.~148 are:

\begin{equation}
\begin{aligned}
z_{2,i}=\sum_{k\in \mathrm{TRIM}}^{{k_{i}=\pi}} \frac{\mathrm{N}_{-}(k)-\mathrm{N}_{+}(k)}{2}\ mod\ 2,\ i=1,2,3,
\end{aligned}
\label{1}
\end{equation} 

\begin{equation}
\begin{aligned}
z_{4}=\sum_{k\in \mathrm{TRIM}} \frac{\mathrm{N}_{-}(k)-\mathrm{N}_{+}(k)}{2}\ mod\ 4,
\end{aligned}
\label{2}
\end{equation} 
where TRIM stands for the time-reversal-invariant momenta, and $\mathrm{N}_{\pm}(k)$ represents for the number of bands that are even/odd under inversion symmetry at $k$.

Following Eq.~(\ref{1}) and (\ref{2}), we obtain a series of topological invariants with $z_{2,1}z_{2,2}z_{2,3}z_{4}$ = (1110), meaning that there are 6 $mod$ 12 nodal lines along the [111] direction~\cite{fu2007topological,fukui2007quantum, zhang2020diagnosis,zhang2021predicting}. This conclusion is confirmed by DFT calculations that show six endless Dirac-nodal lines along the [111] direction (red lines in Fig.~\ref{fig:FIG3} b-c). These Dirac nodal lines carry a $\pi$ Berry phase are protected by the $\mathcal{PT}$ symmetry. In addition to Dirac nodal lines, the DFT calculations reveal two nodal-rings, similar to those observed in the magnetic Weyl semimetal \CSS{}. These nodal-rings have { binding energies about 80 meV higher than those endless ones} and can be understood by decomposing the symmetry-based indicator as  $z_{2,1}z_{2,2}z_{2,3}z_{4}$ = (1110) = (1112) + (0002) {with $z_{4}$ $mod$ 4}, where (1112) and (0002) uncover the 6 $mod$ 12 Dirac-nodal lines along [111] direction and 2 nodal rings located at inversion-related momenta, respectively~\cite{zhang2020diagnosis,zhang2021predicting}. 

\begin{figure*}
\includegraphics[scale=0.4]{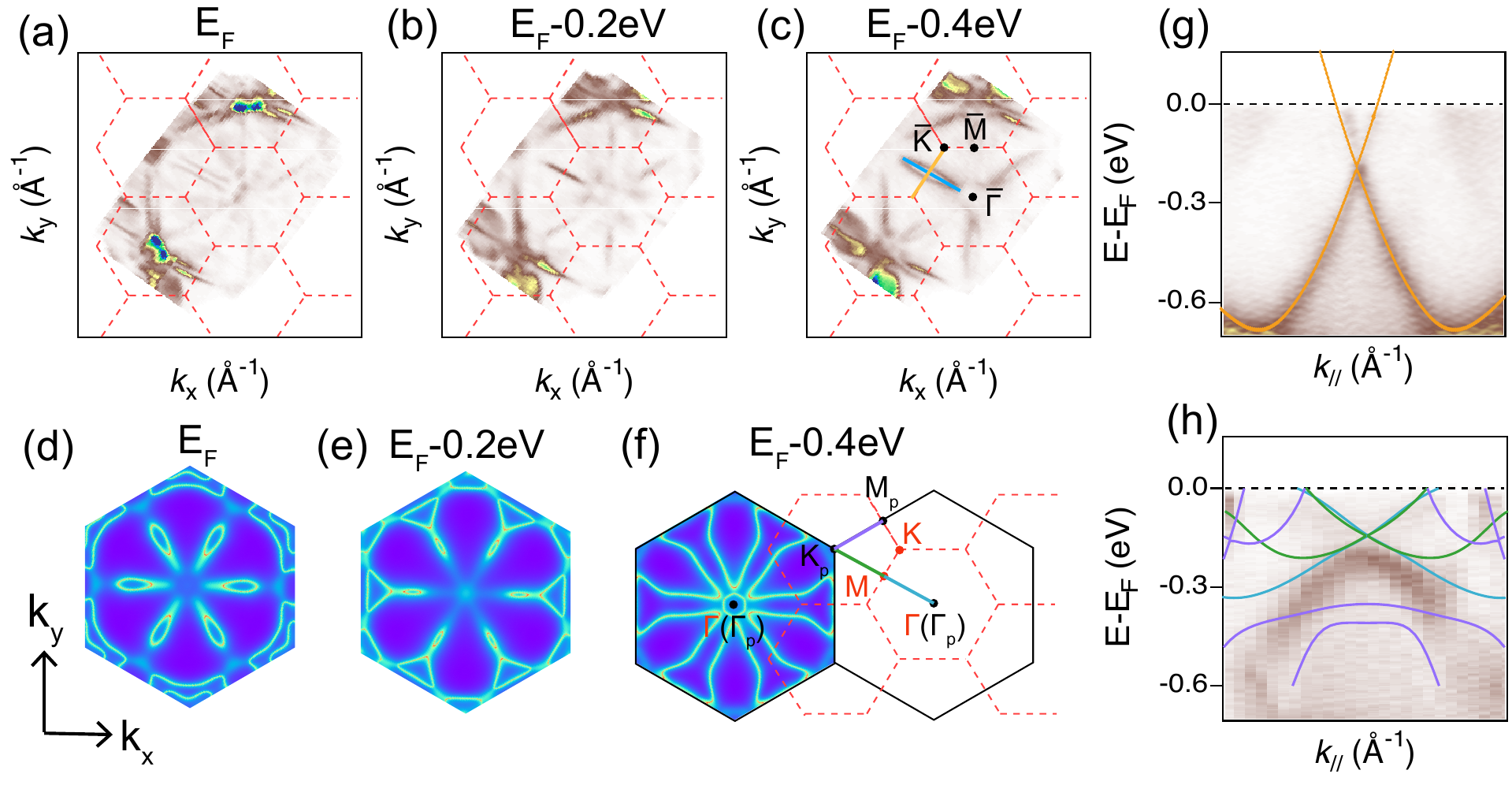}\caption{\textbf{Fermi surface and band dispersion obtained from ARPES measurements and DFT calculations.} \textbf{a-c} Fermi surface mapping with energy of $\textrm{E}_\textrm{{F}}$, $\textrm{E}_\textrm{{F}}-$0.2 eV and $\textrm{E}_\textrm{{F}}-$0.4 eV. $\bar{\Gamma}$ is the center momenta in the first surface BZ. 
\textbf{d-g} DFT calculations on the Fermi surface mapping with $k_{z}$ = $\pi$, which correspond to the energies in \textbf{b-d}. We note that the BZ in \textbf{a-c} are with hexagonal type lattice, i.e., conventional cell of \NIS, which will be three times smaller than the rhombohedral BZ used in the DFT calculation. BZ folding from the DFT calculated one (black hexagon) to the experimental one (red dashed hexagon) is shown in \textbf{f}. \textbf{g-h} are two bulk band cutting marked in \textbf{c}, with $k_{\mathrm{\parallel}}$ follow the yellow and blue lines, respectively.
\label{fig:arpes}}
\end{figure*}

\subsection*{Angle-resolved photoemission spectroscopy measurements}
To study the electronic structure and its implications experimentally, crystals of \NIS{} were grown and characterized as described in Supplementary Note. 1. Single crystal x-ray diffraction confirmed the structure reported in Ref.~\cite{weihrich2005half}. The crystals are metallic with residual resistivity ratios near 130 K and showed Pauli paramagnetic behavior with no evidence of phase transitions between 2 K and 300 K. To identify the theoretically predicted Dirac nodal lines, we performed ARPES measurements on the (001) surface of \NIS{} crystals at 10~K. Figure~\ref{fig:arpes} a-c show the ARPES constant-energy plot with binding energy, $\mathrm{E}_\mathrm{B}$=0, 0.2 and 0.4~eV, respectively. The red dashed hexagon-network indicates the surface BZ. The data was collected at the photon energy, $h\nu$=125~eV, corresponding to $k_{z}\sim0$ (see Supplementary Figure. 2).  The Fermi surface near the surface high-symmetry $\bar{\mathrm{M}}$ point shows parallel lines connecting the first and second BZ. Moving to high binding energy, the parallel lines first evolve to a node near $\mathrm{E}_\mathrm{B}$=0.2~eV and eventually turn into an oval centered at the $M$ point. These constant-energy plots suggest a Dirac point at the $\bar{\mathrm{M}}$ point and is qualitatively consistent with the bulk electronic structure shown in Fig.~\ref{fig:arpes} d-f. We emphasize that DFT calculations use the primitive cell convention with a bigger BZ size compared with the surface BZ. Therefore the calculated electronic structure must be folded to the surface BZ, as shown in Fig.~\ref{fig:arpes} f. 
Figures~\ref{fig:arpes} g and h show the ARPES intensity plot along the surface $\bar{\mathrm{M}}-\bar{\mathrm{K}}$ and $\bar{\Gamma}-\bar{\mathrm{M}}$ direction, respectively. The Dirac-cone structure is clearly resolved along the $\bar{\mathrm{M}}-\bar{\mathrm{K}}$-direction (Fig.~\ref{fig:arpes} g) and consistent with the DFT calculations. 
{However the DFT calculated electronic structure shown in Fig. 3h is only partially resolved. We suspect this is due to the ARPES matrix element effect \cite{Wang2012} combined with the zone folding effect. As we show in Fig. 3 f, the surface Brillouin zone is smaller than the bulk Brillouin zone. Consequently, the purple and green bands shown in Fig. 3h are folded from the bulk BZ and hence may show weaker intensity on the surface.}

\subsection*{Transport property measurements}
\begin{figure*}
\includegraphics[scale=0.8]{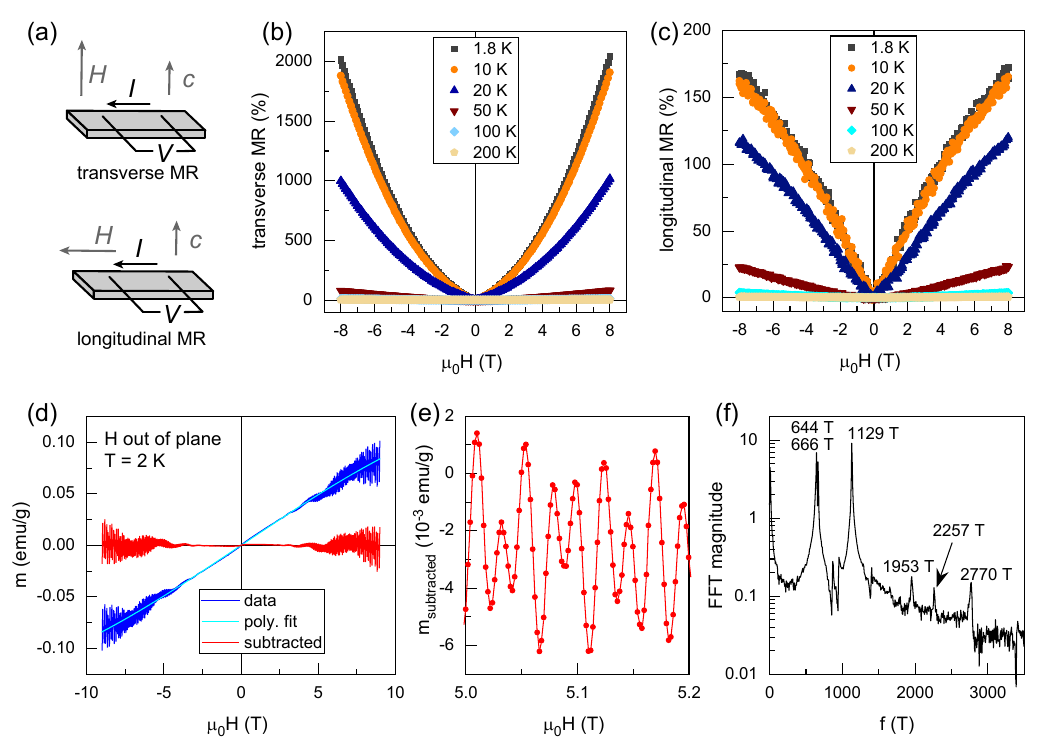}\caption{\textbf{Magnetic field dependence of the resistivity and magnetization of \NIS.} \textbf{a} Relative orientations of the current $I$, magnetic field $H$, and crystallographic $c$-axis for the magnetoresistance measurements. \textbf{b-c} Transverse and longitudinal magnetoresistance MR(\%) = [$\mathrm{R(H)}$-$\mathrm{R}$(0)]/$\mathrm{R}$(0)*100 measured at the indicated temperatures. \textbf{d} Measured magnetization in blue with quantum oscillations clearly seen at higher fields. The oscillating part of the signal (red) is determined by subtracting a polynomial fit (cyan) to the measured data, and is shown over a small field range in \textbf{e}. \textbf{f} Results of Fourier transforming of the oscillating component vs 1/B.  
}
\label{fig:mr}
\end{figure*}

The observation of Dirac nodal lines near the Fermi level is expect to affect transport properties of \NIS{}. The small effective mass of the Dirac band is expected to significantly enhance the carrier mobility and yield giant magnetoresistance \cite{Song2015, ZhangJ2021, Ok2021}. To confirm these effects, in Figs.~\ref{fig:mr} b-c we show the transverse and longitudinal magnetoresistance MR(\%) = 100[$\mathrm{R(H)}$-$\mathrm{R}$(0)]/$\mathrm{R}$(0) measured at different temperatures indicated in the figures. The measurement orientations are shown in Fig.~\ref{fig:mr} a. 
Indeed, we find that the longitudinal magnetoresistance at 1.8 K is non saturating up to 9 T and reaches value of 2000\%, supporting a giant magnetoresistance induced by the Dirac nodal line. In the longitudinal geometry, the magnetoresistance is more than one order of magnitude smaller than that in the transverse geometry, consistent with the quasi-two-dimensional electronic structure. Similar magnetoresistance behavior in another crystal is shown in Supplementary Figure. 6.

High carrier mobility is also evidenced by the clear quantum oscillations seen in the magnetization data in Fig.~\ref{fig:mr} d. The presence of several frequencies in these de Haas-van Alphen oscillations is apparent in the small field range shown in Fig.~\ref{fig:mr} e. Fourier transformation of the oscillating component vs 1/$\mathrm{B}$ shown in Fig.~\ref{fig:mr} f gives components corresponding to diameters of $k_\mathrm{F}=$0.12, 0.15, 0.2, 0.22 and 0.247 in $\frac{\pi}{a}$ ($a$ is lattice constant) for the Fermi surface on $k_z=$0 plane, in agreement with the DFT calculation (see Supplementary Figure. 9).


\section*{Discussion}
Our theoretical and experimental results establish \NIS{} as the first nonmagnetic Kagome material hosting endless Dirac nodal lines near the $\textrm{E}_\textrm{{F}}$. \NIS{} is therefore a promising material platform to engineer diverse topological electronic states. In the atomic limit, Ni has an even number of electrons, making \NIS{} an ``insulator''. In the presence of spin-orbit coupling, \NIS{} is actually a weak topological insulator with topological invariants of $z_{2,1}z_{2,2}z_{2,3}z_{4}$=(1112). This indicates that there will be an even number of 2D Dirac cones. Indeed, surface calculations on the (001) surface shows 4 Dirac cones locating at each of the time-reversal-invariant momenta on the surface BZ (see 
Supplementary Note. 3). \NIS{} may also turn into a strong topological insulator by making band inversions at the high-symmetry $\mathrm{T}$-point through, $e.g.$, external strain. Furthermore, the inversion and/or time-reversal symmetry-breaking will lift the band degeneracy and induce Weyl points in \NIS{}, {such as doping of cobalt elements, which may induce a ferromagnetic phase transition and lead to a strong enhancement of physical properties like Nernst conductivity \cite{yanagi2021first,irkhin2021topological}}. 
Finally, although multiple bands are crossing $\textrm{E}_\textrm{{F}}$, previous studies of Dirac/Weyl semimetals \cite{Yin2020, Ok2021} have shown that the transport properties are determined by massless electrons near the Fermi level. 
As we show in Fig.~\ref{fig:arpes}, the experimentally determined Dirac-point is about 0.2~eV below $\textrm{E}_\textrm{{F}}$ and possibly responsible for the giant magnetoresistance in \NIS{}. Our DFT calculation shows that 0.65~$e$ per unit cell is sufficient to bring the Dirac cone to the Fermi level for quantum transport behaviors \cite{Ok2021}. This can possibly be realized by substituting Ni with $\sim$20\% Co or substituting In with $\sim$30\% Sn. 
{Since In atom is nonmagnetic and contribute only small density of state near the fermi level, Sn substitution of In may induce less disorder effects. The magnetic Co substitution may, however, break the time-reversal-symmetry and give rise to Weyl points as those observed in Co$_3$Sn$_2$S$_2$ \cite{xu2018topological,LiuDF2019,morali2019fermi,jiao2019signatures,shen2019anisotropies,chen2019pressure,ding2019intrinsic,xu2020electronic,li2020epitaxial,tanaka2020topological,gopal2020observation,muechler2020emerging,guin2019zero}.}

In summary, we observed endless Dirac nodal lines in Kagome metal \NIS{}. Our results reveal a diverse topological landscape of multi-band \kag{} metals and suggest \NIS{} as a promising material platform to engeering topological electronic structures.

\section*{Methods}
We performed first-principle calculations based on density functional theory \cite{hohenberg1964inhomogeneous} within the Perdew-Burke-Ernzerhof exchange-correlation \cite{perdew1996generalized} using the Vienna $ab$ $initio$ Simulation Package (VASP) \cite{kresse1996efficient}. The plane-wave cutoff energy is set to be 400 eV with a 11$\times$11$\times$11 $k$-mesh in the BZ for the self-consistent calculations, and all the calculations are made considering absence of SOC. 
Fermi surface and surface state calculations are performed using the tight-binding model of Ni$_{3}$In$_{2}$S$_{2}$, which is obtained from maximally localized Wannier functions \cite{marzari1997maximally}. 
The lattice constants used in our calculation are $a$ = $b$ = 5.37 $\mathrm{\AA}$ and $c$ = 13.56 $\mathrm{\AA}$, which match with the experimental values of the atomic sites and the lattice constants.
{For the calculation with hole doping, we reduce the number of total valence electron by 0.65 $e$ per unit cell and compensate it by a uniform background of positive charges via VASP.}

The ARPES experiments were performed on \NIS{} single crystals. The samples were cleaved $in$-$situ$ in a vaccum better than 5$\times$10$^{-11}\  torr$. The experiment was performed at beamline 21-ID-1 at National Synchrotron Light Source II, Brookhaven. The measurements were taken with synchrotron light source and a Scienta-Omicron DA30 electron analyzer. The total energy resolution of the ARPES measurement is $\sim$15~meV. The sample stage was maintained at low temperature ($T=$15~K) throughout the experiment.

Transport and heat capacity data were collected using a Quantum Design Dyancool cryostat (see 
Supplementary Note. 2). Isothermal magnetization curves were also measured in this cryostat using the vibrating sample magnetometer option, while the temperature dependence of the magnetic susceptibility was measured using a Quantum Design MPMS3 usign the DC measurement option. Contacts for transport measurements were made using Epotek H20E silver epoxy.

\section*{DATA AVAILABILITY}
The datasets generated during and/or analysed during the current study are available from the corresponding authors on reasonable request.

\section*{CODE AVAILABILITY}
The related codes are available from the corresponding authors on reasonable request.

\section*{Acknowledgements}
This research at Oak Ridge National Laboratory (ORNL) was sponsored by the U.S. Department of Energy, Office of Science, Basic Energy Sciences, Materials Sciences and Engineering Division (ARPES experiment crystal growth, and physical properties measurements). T. Z. and S. M. acknowledge the supports from Tokodai Institute for Element Strategy (TIES) funded by MEXT Elements Strategy Initiative to Form Core Research Center Grants Nos. JPMXP0112101001, JP18J23289, JP18H03678, and JP22H00108. 
T. Z. also acknowledge the support by Japan Society for the Promotion of Science (JSPS), kAkENHI Grant No. 21k13865. ARPES measurements used resources at 21-ID-1 beamlines of the National Synchrotron Light Source II, a US Department of Energy Office of Science User Facility operated for the DOE Office of Science by Brookhaven National Laboratory under contract no. DE-SC0012704.

\section*{AUTHOR CONTRIBUTIONS}
M.A.M, H.M. and T.Z. devised the project idea and prepared the manuscript. 
T.Z. and S.M. performed the first-principles calculations. 
T.Y., E.V., H.L., R.G.M., H.N.L, H.M and M.A.M prepared the material sample, performed the experiments and analyzed the data. 
All the authors discussed the results and the ideas for analysis, and approved the complete version.

\section*{COMPETING INTERESTS}
The authors declare no competing interests. 
This manuscript has been authored by UT-Battelle, LLC under Contract No. DE-AC05-00OR22725 with the U.S. Department of Energy. The United States Government retains and the publisher, by accepting the article for publication, acknowledges that the United States Government retains a non-exclusive, paid-up, irrevocable, world-wide license to publish or reproduce the published form of this manuscript, or allow others to do so, for United States Government purposes. The Department of Energy will provide public access to these results of federally sponsored research in accordance with the DOE Public Access Plan (http://energy.gov/downloads/doe-public-access-plan).

\bibliographystyle{naturemag}

\bibliography{newref}

\end{document}